\documentclass[12pt]{article}
\input{epsf.tex}
\def\ba{\begin{eqnarray}}
\def\ea{\end{eqnarray}}
\def\psit{\stackrel{-}{\psi}}
\def\a{\alpha}
\def\b{\beta}
\def\s{\sigma}
\def\r{\rho}
\def\d{\delta}
\def\e{\varepsilon}
\def\g{\gamma}
\def\part{\partial}
\def\parlr#1{\stackrel{\leftrightarrow }{\partial #1}}
\def\parr#1{\stackrel{\rightarrow }{\partial #1}}
\def\parl#1{\stackrel{\leftarrow }{\partial #1}}
\begin{document}
\title{\Large \bf 
Static polarizability vertex and its applications
}
\author{A.~Ilyichev
$^{a}$\thanks{E-mail: ily@hep.by}, 
S.~Lukashevich$^{b}$\thanks{E-mail: lukashevich@gsu.unibel.by}, 
N.~Maksimenko$^{b}$\thanks{E-mail: maksimenko@gsu.unibel.by}\\
{\small \it $^a $National Scientific	
and Educational Centre of Particle and High Energy Physics}
\\[-1mm]
{\small \it  of the Belarusian State University,
220040  Minsk,  Belarus}\\
{\small \it $^b $Gomel State  University, 246699 Gomel, Belarus}}
\date{}
\maketitle

\vspace*{-10mm}
{\abstract
Using the Lagrangian that was developed on
the corresponding principle between the moving medium electrodynamic and quantum field theory
the explicit expression for the static polarizability vertex 
has been obtained.  The applications of this vertex
for calculations of real Compton scattering amplitude 
as well as the imaginary part of doubly virtual
Compton scattering amplitude have been demonstrated.}
\section{Introduction}
Nowadays a set of the low energy experiments are performed to investigate
the  nucleon structure within non-perturbative QCD-region.
Some of them allows ones not only to investigate 
the nucleon elastic properties
at low $Q^2$-region but also to see its  internal structure 
 by the  different polarizabilities measurement
\cite{sp}.

Performing a such kind of the experiments it will be also important to 
have some theoretical explanation of the measured quantities.  One of 
the models that can be used
for the estimation of $Q^2$-dependence of the nucleon spin polarizabilities is based on
the corresponding principle between the moving medium electrodynamic and quantum field theory.
Nowadays the Lagrangian describing this theory was constructed, presented
in \cite{Cr} and for the photon-nucleon tensor interaction has a form
\ba
\label{lag}
{\cal L}^{pol}_{eff}=-\frac {i\pi}{M} \left(\psit
 {\g^\mu}\parlr{ _\nu} \psi
\right)
(\a_0 F_{\mu \rho} F^{\rho \nu} - \b _0 {\widetilde F}_{\mu \rho } {\widetilde F}^{\rho \nu }).
\ea
Here $~~\parlr { _\nu}=\parl { _\nu}-\parr { _\nu}$, $\psi$ is the wave functions
of the nucleon whose mass is $M$, 
$F_{\mu \nu}=\partial_{\mu}A_{\nu }-\partial_{\nu}A_{\mu }$ is
electromagnetic tensor,
${\widetilde F}_{\mu \nu}=\e_{\mu \nu \rho \sigma} F^{\rho \sigma}/2$. 
Presented above Lagrangian depends on the static electric 
$\alpha _0$ and magnetic  $\beta _0$ polarizabilities
whose numerical value are known (see \cite{sp,bab} and references
therein).

Having Lagrangian that described any interaction between particles we can
use the standard Feynman rule technic. 
To apply presented above Lagrangian for this method
in the present report we define the vertex of interaction
and consider its application for calculation of real Compton scattering (RCS)
 and doubly virtual
Compton scattering (VVCS). 

\section{Static polarizability vertex}
In order to obtain the explicit expression for the polarizability vertex 
we will follow to the notations of Appendix B of \cite{chl}.
According to this approach our Lagrangian (\ref{lag})
can be presented as: 
\ba
{\cal L}^{pol}_{eff}=
\int \prod_{i=1}^4
\left[ d^4x_i\delta^4(x-x_i) \right ]
\alpha _{\s \d}^{ r'r}(x_1,x_2,x_3,x_4)
\psit_{r'}(x_3)\psi_{r}(x_1)
A^{\s }(x_4)A^{\d }(x_2),
\ea
where $r'r$ are the four-spinor indexes (that usually dropped).
Taking into account that
\ba
\widetilde {F}_{\mu\rho} \widetilde {F}^{\rho \nu}=
F_{\mu\rho} F^{\rho \nu}+\frac 1 2\d_{\mu}^{ \nu}F_{\rho \sigma} F^{\rho \sigma}
\ea
we can immediately find that
\ba
\alpha _{\s \d}^{ r'r}(x_1,x_2,x_3,x_4)&=&-\frac {i\pi}{M}
\left(
\frac{\part}{\part x_{3\nu}}-
\frac{\part}{\part x_{1\nu}}
\right)\gamma_{\mu}^{ r'r}
\left((\a_0-\b_0)
\left[
\frac{\part}{\part x_{4\mu}}\d^{\r}_{ \s}
-\frac{\part}{\part x_{4\r}}\d^{\mu}_{ \s}
\right]
\right .\nonumber \\&& \left . \times
\left[
\frac{\part}{\part x_{2}^{\r}}g_{\nu \d}
-\frac{\part}{\part x_{2}^{\nu}}g_{\r \d}
\right]
-\frac 12 \d^{\mu}_{ \nu}\b_0
\left[
\frac{\part}{\part x_{4\r}}\d^{\g }_{ \s}
-\frac{\part}{\part x_{4\g}}\d^{\r }_{ \s}
\right]
\right .\nonumber \\&& \left . \times
\left[
\frac{\part}{\part x_{2}^{\r}}g_{\g \d}
-\frac{\part}{\part x_{2}^{\g}}g_{\r \d}
\right]
\right)
\nonumber \\&=&\int
\frac{d^4p_1}{(2\pi)^4}
\frac{d^4p_2}{(2\pi)^4}
\frac{d^4q_1}{(2\pi)^4}
\frac{d^4q_2}{(2\pi)^4}
{\tilde \alpha }_{\s \d}^{ r'r}(p_1,q_1,p_2,q_2)
\times
\nonumber \\&&
\times
e^{ip_1(x-x_1)+iq_1(x-x_2)-ip_2(x-x_3)-iq_2(x-x_4)},
\ea
\begin{figure}[t!]
\vspace*{3mm}
\unitlength 1mm
\hspace*{2cm}
\begin{picture}(80,80)
\put(10,0){
\epsfxsize=12cm
\epsfysize=12cm
\epsfbox{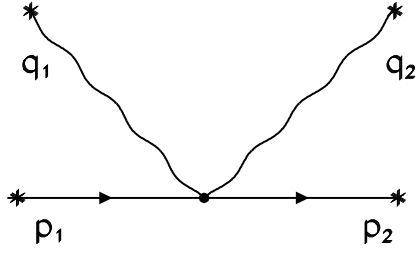}
}
\end{picture}
\vspace*{-50mm}
\caption{
\label{pp}
Polarizability vertex
}
\end{figure}
\noindent
where $p_1$ and $q_1$ ($p_2$ and $q_2$)
are the incoming (outgoing) nucleon and photon 
momenta respectively (see Fig.~\ref{pp}).
As a result 
${\tilde \alpha }_{\s \d}^{ r'r}(p_1,q_1,p_2,q_2)$
can be found  by the following substitution: 
\ba
{\tilde \alpha }_{\s \d}^{ r'r}(p_1,q_1,p_2,q_2)
&=&\alpha _{\s \d}^{ r'r}(x_1,x_2,x_3,x_4)
\left [
\frac{\part}{\part x_{1}}\to ip_1,\;
\frac{\part}{\part x_{2}}\to iq_1,\;
\frac{\part}{\part x_{3}}\to -ip_2,\;
\frac{\part}{\part x_{4}}\to -iq_2
\right]
\nonumber \\
&=&-\frac {\pi}{M}
(p_{1}^{\nu}+p_{2}^{\nu})\g_{\mu}^{ r'r}
\left((\a_0-\b_0)
\displaystyle
[q_{2}^{\mu}\d^{\r}_{ \s}-q_{2}^{\r}\d^{\mu }_{\s}]
[q_{1\r}g_{\nu \d}-q_{1\nu}g_{\r \d}]
\right .
\nonumber \\&&
\left .
-\frac 12 \d^{\mu }_{ \nu}\b_0
\left[
q_{2}^{\r}\d^{\g }_{\s}
-q_{2}^{\g}\d^{\r}_{ \s}
\right]
\left[
q_{1\r}g_{\g \d}
-q_{1\g}g_{\r \d}
\right]
\right).
\ea
Summing up over the all symmetric states
and multiply the result on $i$
we receive the final expression
for polarizability vertex that presented on Fig.~\ref{pp} 
\ba
\Gamma_{\s \d}^{pol}(p_1,q_1,p_2,q_2)&=&i(
{\tilde \alpha }_{\s \d}(p_1,q_1,p_2,q_2)+
{\tilde \alpha }_{\d \s}(p_1,q_1,p_2,q_2)+
{\tilde \alpha }_{\s \d}(p_1,-q_2,p_2,-q_1)+
\nonumber \\&&\;\;\;
{\tilde \alpha }_{\d \s}(p_1,-q_2,p_2,-q_1)
).
\ea
Here we dropped the four-spinor indexes. 

\section{RCS}
As a simplest application for the presented above vertex let us 
consider RCS 
\ba
\g (k)+N(p)\to\g (k')+N(p') 
\ea
($k^2=k'^2=0$, $p^2=p'^2=M^2$)
that in lowest order are described  by the following matrix element:
\ba
iT&=&\frac 14\e^{\mu *}\e^{\nu}
{\bar u}(p')\Gamma_{\mu \nu}^{pol}(p,k,p',k'){\bar u}(p')=
-i\e^{\mu *}\e^{\nu}
\frac{\pi}M{\bar u}(p')
(2M\b_0(K_{\mu}K_{\nu}-Q_{\mu}Q_{\nu})
\nonumber \\&&
+(\a_0-b_0)[\g \cdot K (P_{\mu }K_{\nu }+P_{\nu }K_{\mu })
-2K^2(\g_{\mu}P_{\nu}+\g_{\nu}P_{\mu})
+K\cdot P(\g_{\mu}K_{\nu}+\g_{\nu}K_{\mu})]
\nonumber \\&&
-2g_{\mu \nu }[2M\b_0 K^2+(\a_0-\b_0)\g \cdot K K\cdot P])u(p),
\ea
where we use the standard notations \cite{bab}
$\displaystyle P=\frac 1 2(p+p')$,
$\displaystyle K=\frac 1 2(k+k')$,
$\displaystyle Q=\frac 1 2(p-p')=\frac 1 2(k'-k)$.
Taking into account
\ba
{\bar u}(p')\g\cdot Pu(p)=M{\bar u}(p')u(p),\;
{\bar u}(p')\g\cdot Qu(p)=0,\;
Q^2=-K^2,\; P\cdot Q=0
\ea
one can see that this result is agree with amplitude obtained in  \cite{Cr}.

\section{VVCS}
Today one of the most interesting topic in 
polarizability investigation is the measurement of
$Q^2$-dependence of 
the forward polarizabilities in deep-inelastic scattering (DIS)
\cite{dd} that can be presented as the imaginary part of
VVCS \cite{sp}.
To shed light on the question about static and dynamic 
polarizability relations we propose 
to express the forward polarizabilities via static ones
applying the standard Feynman rule technic.

\begin{figure}[t!]
\vspace*{3mm}
\begin{tabular}{cc}
\unitlength 1mm
\hspace*{2cm}
\begin{picture}(80,80)
\put(-30,0){
\epsfxsize=12cm
\epsfysize=12cm
\epsfbox{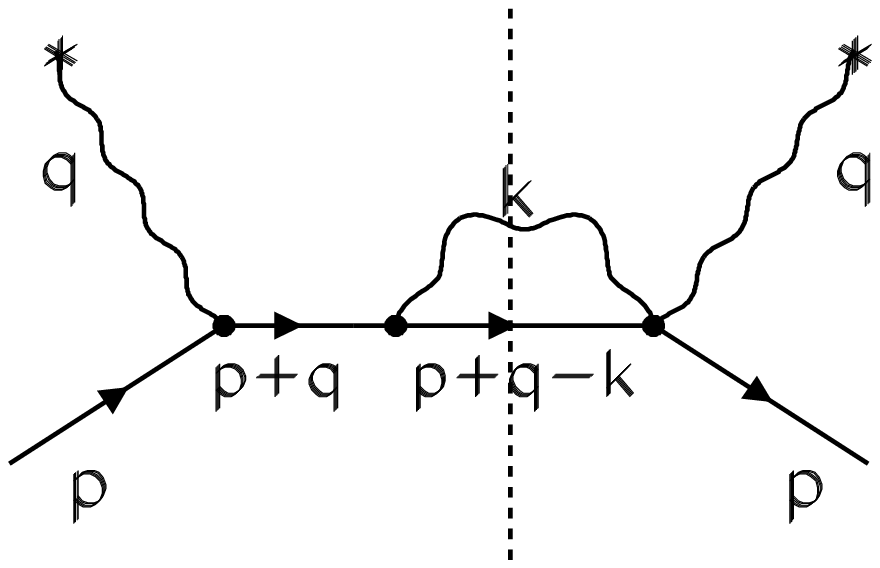}
\put(-65,50){\makebox(0,0){a)}}
}
\end{picture}&
\begin{picture}(80,80)
\put(-130,0){
\epsfxsize=12cm
\epsfysize=12cm
\epsfbox{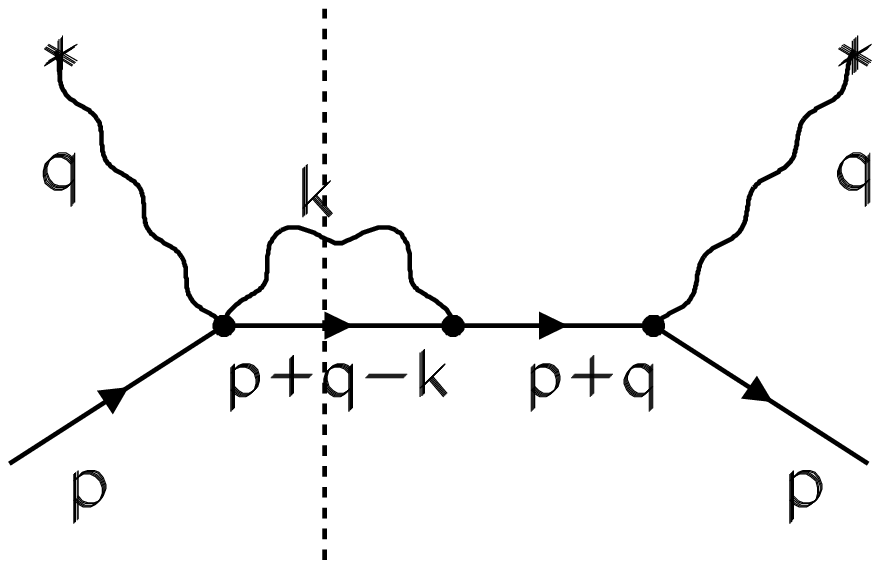}
\put(-180,140){\makebox(0,0){b)}}
}
\end{picture}
\\[15mm]
\hspace*{2cm}
\begin{picture}(80,80)
\put(-160,0){
\epsfxsize=12cm
\epsfysize=12cm
\epsfbox{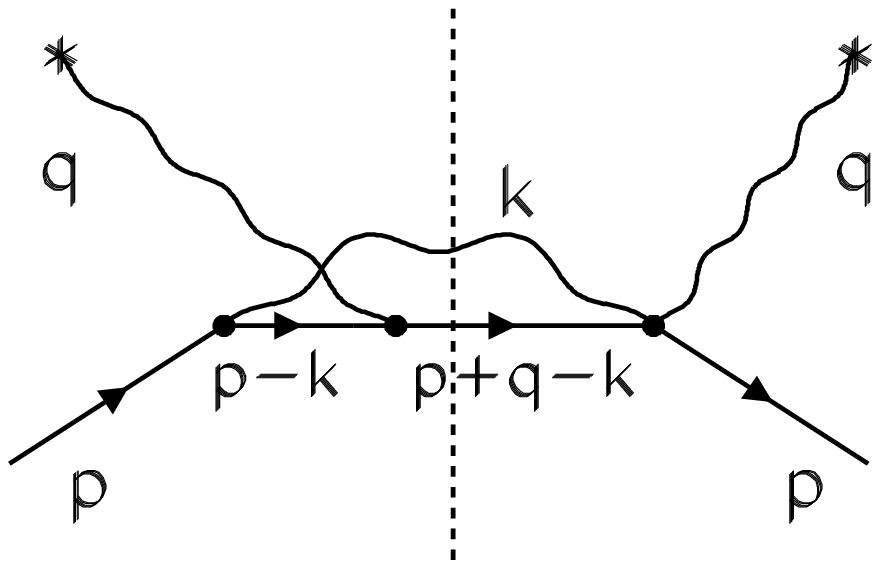}
\put(-175,145){\makebox(0,0){c)}}
}
\end{picture}&
\begin{picture}(80,80)
\put(-130,0){
\epsfxsize=12cm
\epsfysize=12cm
\epsfbox{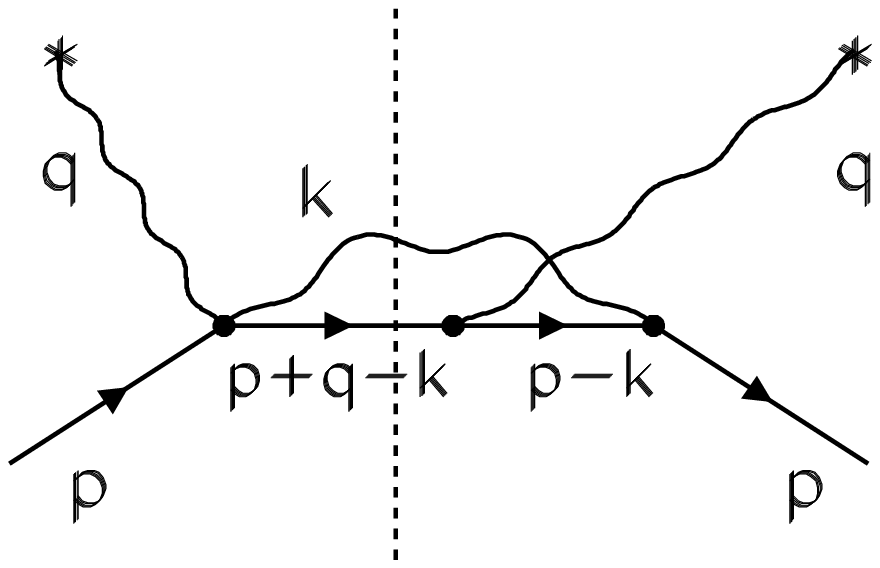}
\put(-180,145){\makebox(0,0){d)}}
}
\end{picture}
\\[13mm]
\hspace*{1cm}
\begin{picture}(80,80)
\put(-50,0){
\epsfxsize=12cm
\epsfysize=12cm
\epsfbox{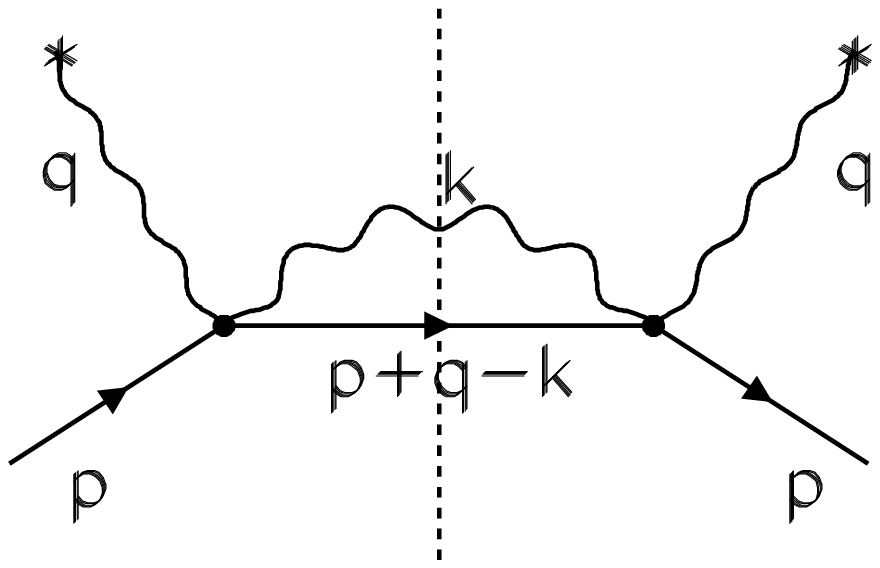}
\put(-180,142){\makebox(0,0){e)}}
}
\end{picture}&

\end{tabular}
\vspace*{-43mm}
\caption{
\label{vvc}
Feynman graphs for the lowest order VVCS amplitudes 
whose imaginary parts give contributions 
to the hadronic tensor for inclusive DIS. The dashed lines show the cuts 
for imaginary part calculations.   
}
\end{figure}
The lowest order VVCS amplitudes with  static polarizabilities to 
whose imaginary part gives non-zero contribution to the hadronic tensor of
inclusive DIS are presented by Feynman grafs on Fig.~\ref{vvc}. 
 It should be
noticed that contribution from Fig.~\ref{vvc}~(e) is negligible
($\sim 10^{-8}$~fm$^6$) and can be dropped. 
Performing cut over dash line on Fig.~\ref{vvc}~(a-d)
the imaginary part of these amplitudes can be presented as 
\ba
{\rm Im \;}T=\pi^2\e^*_{\nu}(q)\e_{\mu}(q)
\int d\Theta
{\bar u}(p)
\left[\frac{
\Gamma^{\mu \nu \;a}_{VVCS}
+\Gamma^{\mu \nu \;b}_{VVCS}
}{(p+q)^2-M^2}+
\frac{
\Gamma^{\mu \nu \;c}_{VVCS}+
\Gamma^{\mu \nu \;d}_{VVCS}
}{(p-k)^2-M^2}\right]
u(p),
\label{im}
\ea
where $d\Theta=d^4k\delta(k^2)\delta((p+q-k)^2-M^2)$,
\ba
\Gamma^{\mu \nu \; a}_{VVCS}&=&
\Gamma^{pol \;  \nu \a}(p+q-k,k,p,q)
({\hat p}+{\hat q}-{\hat k}+M)
\Gamma_{\a}^{el}(-k)
({\hat p}+{\hat q}+M)
\Gamma^{el \; \mu}(q),
\nonumber \\[1mm]
\Gamma^{\mu \nu \; b}_{VVCS}&=&
\Gamma^{el \; \nu}(-q)
({\hat p}+{\hat q}+M)
\Gamma_{\a}^{el}(k)
({\hat p}+{\hat q}-{\hat k}+M)
\Gamma^{pol \; \mu \a}(p,q,p+q-k,k),
\nonumber \\[1mm]
\Gamma^{\mu \nu \; c}_{VVCS}&=&
\Gamma^{pol \; \nu \a}(p+q-k,k,p,q)
({\hat p}+{\hat q}-{\hat k}+M)
\Gamma^{el \; \mu }(q)
({\hat p}-{\hat k}+M)
\Gamma_{\a}^{el}(-k),
\nonumber \\[1mm]
\Gamma^{\mu \nu \; d}_{VVCS}&=&
\Gamma_{\a}^{el}(k)
({\hat p}-{\hat k}+M)
\Gamma^{el \; \nu}(-q)
({\hat p}+{\hat q}-{\hat k}+M)
\Gamma^{pol \;\mu \a}(p,q,p+q-k,k),
\ea
and
$
\Gamma_{\mu}^{el}(q)=-ie
\left (F_D(-q^2)\g_{\mu}+F_P(-q^2)i\s_{\mu \a}q^{\a}/2M\right)
$ 
is the usual elastic vertex.

Notice that according to \cite{sp} 
from the expression (\ref{im}) one can extract the
partial cross sections $K\sigma_T$, $K\sigma_L$, $K\sigma_{TT}$, $K\sigma_{LT}$
and calculate $Q^2$-dependences of forward polarizabilities in a following way:
\ba
\alpha(Q^2)+\beta(Q^2)=
\frac 1{2 \pi ^2}\int \limits_{\nu_0}^{\infty}
\frac {K(Q^2,\nu )}{\nu}
\frac {\sigma_{T}(Q^2,\nu )}{\nu^2}d\nu,\;
\alpha_L(Q^2)=
\frac 1{2 \pi ^2}\int \limits_{\nu_0}^{\infty}
\frac {K(Q^2,\nu )}{\nu}
\frac {\sigma_{L}(Q^2,\nu )}{\nu^2}d\nu,
\nonumber \\
\gamma _0(Q^2)=
\frac 1{2 \pi ^2}\int \limits_{\nu_0}^{\infty}
\frac {K(Q^2,\nu )}{\nu}
\frac {\sigma_{TT}(Q^2,\nu )}{\nu^3}d\nu,\;
\delta _{LT}(Q^2)=
\frac 1{2 \pi ^2}\int \limits_{\nu_0}^{\infty}
\frac {K(Q^2,\nu )}{\nu}
\frac {\sigma_{LT}(Q^2,\nu )}{Q\nu^2}d\nu.
\ea
In these expressions
 $Q^2=-q^2$, $\nu=E-E'$ and $E$ ($E'$) is an incoming (outgoing) electron energy 
in lab. frame in DIS.

\vspace*{5mm}
{\bf Acknowledgments.}  
The authors would like to thank
Andrei Afanasev and Jian-Ping Chen for stimulating discussions.

 \end{document}